\documentclass[aps,prl,reprint,twocolumn,showpacs,floatfix,superscriptaddress]{revtex4-2}

\usepackage{amssymb,amsmath,amstext}                
\usepackage{graphicx}                                               
\usepackage{epstopdf}                                               
\usepackage{color}                                                     
\usepackage{bm}                                                        
\usepackage{appendix}                                              
\usepackage[utf8]{inputenc}
\usepackage{latexsym}
\usepackage{xcolor}
\usepackage{braket}
\usepackage{ulem}
\usepackage{siunitx}
\usepackage{umoline}
\usepackage{epsfig}
\usepackage{graphics}
\usepackage{psfrag}
\usepackage{mathtools}
\usepackage[english]{babel}
\usepackage[T1]{fontenc}
\usepackage{lmodern}
\usepackage{rotating}
\usepackage{multirow}

\definecolor{lblue} {RGB}{51,71,158}
\usepackage[colorlinks=true,citecolor=blue,linkcolor=blue,urlcolor=lblue]{hyperref}

\begin{document}

\title{Nonstabilizerness dynamics in many-body localized systems}

\author{Pedro R. Nic\'acio Falc\~ao}
\email{pedro.nicaciofalcao@doctoral.uj.edu.pl}
\affiliation{Szkoła Doktorska Nauk \'Scis\l{}ych i Przyrodniczych, Uniwersytet Jagiello\'nski,  \L{}ojasiewicza 11, PL-30-348 Krak\'ow, Poland}
\affiliation{Instytut Fizyki Teoretycznej, Wydzia\l{} Fizyki, Astronomii i Informatyki Stosowanej,
Uniwersytet Jagiello\'nski,  \L{}ojasiewicza 11, PL-30-348 Krak\'ow, Poland}
\author{Piotr Sierant}
\email{piotr.sierant@bsc.es}
\affiliation{Barcelona Supercomputing Center, Barcelona 08034, Spain}
\author{Jakub Zakrzewski}
\email{jakub.zakrzewski@uj.edu.pl}
\affiliation{Instytut Fizyki Teoretycznej, Wydzia\l{} Fizyki, Astronomii i Informatyki Stosowanej,
Uniwersytet Jagiello\'nski,  \L{}ojasiewicza 11, PL-30-348 Krak\'ow, Poland}
\affiliation{Mark Kac Complex Systems Research Center, Uniwersytet Jagiello{\'n}ski, Krak{\'o}w, Poland}
\author{Emanuele Tirrito}
\email{etirrito@sissa.it}
\affiliation{The Abdus Salam International Centre for Theoretical Physics (ICTP), Strada Costiera 11, 34151 Trieste, Italy}
\affiliation{Dipartimento di Fisica ``E. Pancini", Universit\`a di Napoli ``Federico II'', Monte S. Angelo, 80126 Napoli, Italy}
\date{\today}

\begin{abstract}
Nonstabilizerness, also known as ``magic'', quantifies the deviation of quantum states from stabilizer states, capturing the complexity necessary for quantum computational advantage. In this study, we investigate the dynamics of nonstabilizerness in disordered many-body localized (MBL) systems using the stabilizer Rényi entropy (SRE). Leveraging a phenomenological description based on the $\ell$-bit model, we analytically and numerically demonstrate that interactions profoundly influence nonstabilizerness spreading, inducing a power-law growth of SRE that markedly contrasts with the rapid saturation observed in ergodic systems. We validate our theoretical predictions through numerical simulations of the disordered transverse-field Ising model, showing excellent agreement across various disorder strengths, system sizes, and initial states. Additionally, we uncover a universal relationship between SRE and entanglement entropy, revealing their common scaling in the MBL regime independent of disorder strength and system size. Our results offer critical insights into the interplay of disorder, interactions, and complexity in quantum many-body systems.
\end{abstract}

\maketitle

\paragraph*{Introduction. }

Quantum state $\ket{\psi}$ of $L$ qubits is specified by a state vector in $2^L$ dimensional Hilbert space~\cite{nielsen00}. The \textit{exponential} growth of many-body Hilbert space implies that quantum states may become intractable for classical computers for sufficiently large $L$~\cite{40years}, motivating the development of quantum simulators~\cite{Lewenstein07, Cirac12, Altman21} and quantum computers~\cite{DiVincenzo00, Montanaro16, Cerezo21, Bharti22}. However, certain quantum states possess a structure that enables their efficient representation on classical computers. For instance, when $\ket{\psi}$ is weakly entangled~\cite{Vidal03}, it can be simulated at a cost increasing \textit{polynomially} with $L$ using tensor network approaches~\cite{Verstraete08, Schuch08, Schollwoeck11, Haegeman16, Orus19, Banuls23}. Hence, extensive entanglement~\cite{Amico08, Horodecki09} is necessary for quantum devices to reach computational advantage over classical computers~\cite{Preskill12}. Nevertheless, stabilizer states~\cite{Gottesman1998, Gottesman98theory}, may host extensive entanglement, and still be simulated with classical resources scaling \textit{polynomially} with $L$~\cite{Aaronson04, stim}. Therefore, nonstabilizerness, commonly referred to as ``magic'', quantifies the extent to which $\ket{\psi}$  departs from the set of stabilizer states~\cite{Bravyi05, Gross06, Veitch14, Liu22m, Leone2022stabilizer, Leone2024monotone}, and is a quantum resource~\cite{Chitambar19, Turkeshi24makeMagic} essential for characterizing the complexity of quantum states.

Understanding the generation of magic resources in many-body systems is fundamental for assessing their classical simulability. Generically, quantum many-body systems prepared in an out-of-equilibrium state are expected to follow the eigenstate thermalization hypothesis~(\textbf{ETH})~\cite{Deutsch91, Srednicki94, Rigol08, Dalessio16} and to thermalize reaching an equilibrium state described by appropriate ensembles of statistical mechanics~\cite{Foini19, Pappalardi22, Pappalardi23, pappalardi2023microcanonical}. Thermalization is accompanied by fast, ballistic~\cite{Lauchli08, Kim13} or sub-ballistic~\cite{Rakovszky19, Znidaric20}, growth of entanglement entropy, and a rapid saturation of nonstabilizerness measures~\cite{Turkeshi25spreading, Tirrito24anti} to their maximal values~\cite{Turkeshi23a}. The process of thermalization slows down in the presence of disorder~\cite{Luitz16b, Bera17, Sels23dilute, Evers23Internal}. Sufficiently strong disorder leads to a phenomenon of many-body localization (\textbf{MBL})~\cite{Nandkishore15,  Alet18,  Abanin19, Sierant25rev} which prevents the thermalization~\cite{Oganesyan07, Pal10, Huse14, Luitz15, Wahl17, Mierzejewski18, Thomson18, Suntajs20e, Sierant20t, Panda20, Sels20obstruction, Sierant20p, abanin21d, kiefer2021unlimited, Ghosh22} at any experimentally relevant time scale~\cite{Sierant22c, Morningstar22}. The absence of thermalization starkly affects the dynamics of MBL systems, leading to a logarithmic in time growth of entanglement entropy~\cite{DeChiara06, Znidaric08, Bardarson12, Serbyn13a, Iemini16signatures} and the memory of the initial state due to the presence of an emergent set of local integrals of motion, dubbed localized bits or $\ell$-bits~\cite{Serbyn13b, Huse14}. This raises the question: how does nonstabilizerness spread in MBL systems?

\begin{figure}[h!]
    \centering
    \includegraphics[width=1.0\linewidth]{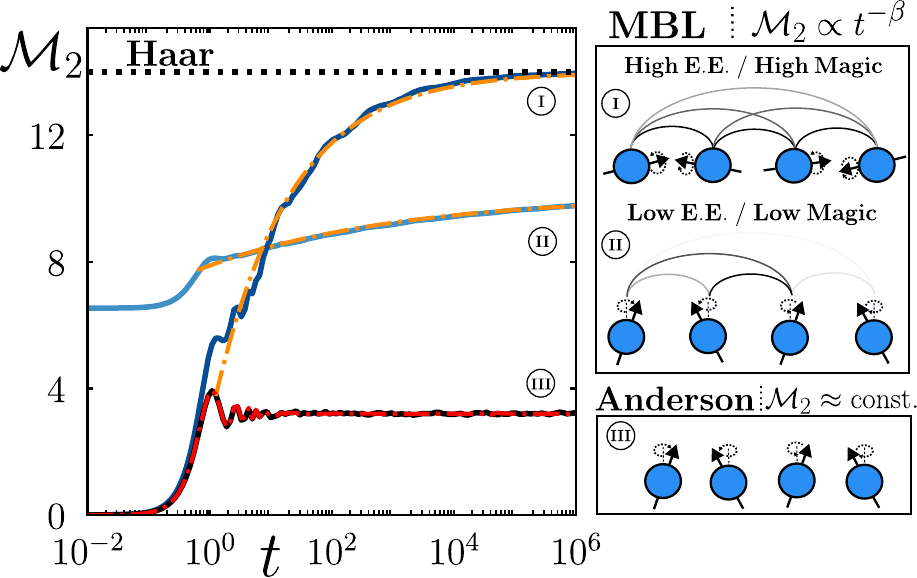}
    \caption{
    {\it Nonstabilizerness spreading in disordered quantum systems.} In the non-interacting case (III, red dashed line), $\mathcal{M}_2$ 
    is analytically tractable (see~\cite{supmat}). It saturates rapidly to a finite value due to the absence of spin dephasing. 
    In the MBL regime, dephasing induces a power-law growth of $\mathcal{M}_2$ toward a saturation value described by Eq.~\eqref{Eq:SRE_MBL}. For the initial X-polarized state (I, orange dashed line), nonstabilizerness grows rapidly and saturates to the Haar value. For a generic product state (II, orange dashed line), the growth is slower and saturates at a lower value, revealing the dependence of nonstabilizerness spreading on the choice of initial state. 
    }
    \label{Fig:lbit}
\end{figure}

In this work, we address this question by investigating the dynamics of nonstabilizerness in strongly disordered spin chains that exhibit MBL. Leveraging the $\ell$-bits, we develop a theoretical framework that describes how nonstabilizerness evolves under MBL dynamics. Specifically, we analyze the stabilizer Rényi entropy (SRE)~\cite{Leone2022stabilizer} as a measure of nonstabilizerness, characterizing its growth in both non-interacting and interacting disorder localized systems. In non-interacting localized systems, the SRE remains limited, while in MBL systems, the slow spin dephasing leads to a power-law growth of nonstabilizerness before its eventual saturation. To corroborate our theoretical predictions, we perform numerical simulations on a microscopic MBL model, the disordered transverse-field Ising model (TFIM). We track the evolution of nonstabilizerness across different disorder strengths, initial states, and system sizes, showing remarkable agreement with the $\ell$-bits-based predictions.  
Additionally, we uncover a universal relationship between the stabilizer Rényi entropy and entanglement entropy, highlighting that their dynamics collapse onto a single curve in the MBL regime, independent of disorder strength and system size.

\paragraph*{Quantifying nonstabilizerness.} 

As a measure of nonstabilizerness, we consider stabilizer R\'{e}nyi entropy (SRE)~\cite{Leone2022stabilizer}, which quantifies the spread of a state in the basis of Pauli string operators, and is defined as

\begin{equation} \label{eq:sre1}
\mathcal{M}_k(|\Psi\rangle) = \frac{1}{1-k}\log_2\left[\sum_{P\in \mathcal{P}_{L}}\frac{\langle \Psi|P|\Psi \rangle^{2k}}{D}  \right],
\end{equation}
where $L$ is the number of qubits, $D=2^L$ is the Hilbert space dimension, $k$ is the R\'enyi index, and $P$ is a Pauli string that belongs to the Pauli group $\mathcal{P}_L$. In particular, $\mathcal{M}_1$ is defined by the limit $k\to 1$ in~\eqref{eq:sre1}, and $\mathcal{M}_{k}\ge 0$, with the equality holding if and only if $|\Psi\rangle$ is a stabilizer state \cite{haug2023stabilizerentropiesand,gross2021schurweylduality}. In our study, we fix $k=2$ and evaluate the SRE using the algorithm of~\cite{sieranttoappear}, which allowed us to obtain numerically exact results for $L\leq 22$. One advantage of the SRE over many other proposed measures of nonstabilizerness~\cite{Veitch14} is that it allows an efficient computation even for a large $L$ ~\cite{PhysRevB.107.035148,tarabunga23m,lami23a,tarabunga24a,Frau24a,Tirrito24a,PhysRevB.111.085144}. 
Moreover, $\mathcal{M}_k(|\Psi\rangle)$ is also experimentally measurable~\cite{Tirrito24a,Turkeshi23b, oliviero22mea,haug23s,niroula24p}.

Quantifying nonstabilizerness and understanding how these resources grow is a current topic of interest in quantum many-body physics~\cite{Liu22m}. Recent works have addressed this question for ergodic many-body systems~\cite{Turkeshi25spreading,Turkeshi23a, dowling2024magic, dowling2025bridging, Tirrito24anti,odavic24s}, where initial state information is rapidly lost and $|\Psi\rangle$ behaves similarly to a random vector~\cite{Dalessio16}. For example, in random unitary circuits, the SRE saturates to the Haar-random state value~\cite{Turkeshi23a}
\begin{equation}\label{Eq:SREH}
    \mathcal{M}_2^{\mathrm{Haar}} = \log_2 (D + 3) - 2,
\end{equation}
at times scaling \textit{logarithmically} with system size $L$ \cite{Turkeshi25spreading}. Generic many-body systems exhibit more intricate behavior. Floquet systems behave similarly to random circuits, while for Hamiltonian dynamics, the time required to approach the Haar value scales linearly in $L$, and the SRE may not reach the Haar value~\cite{odavic24s,Tirrito24anti}.

\paragraph*{The $\ell$-bit model.} 
A characteristic hallmark of MBL is the emergent integrability that microscopic models acquire at sufficiently strong disorder~\cite{Serbyn13b,Huse14,Ros15, Imbrie17}. In the MBL regime, the system is described by a set of $\ell$-bits,  $\hat{\tau}^z_i$, and its Hamiltonian reads
\begin{equation}
    \hat{\mathcal{H}}_{\mathrm{\ell-bit}}= \sum_{i} h_i\hat{\tau}_i^{z} + \sum_{i<j}J_{ij}\hat{\tau}_i^{z}\hat{\tau}_j^{z} + \sum_{i<j<k}J_{ijk}\hat{\tau}_i^{z}\hat{\tau}_j^{z}\hat{\tau}_k^{z} + ...
    \label{eq:l-bit}
\end{equation}
where $h_i$ are random on-site fields drawn uniformly from $[-W,W]$, $J_{ij...}$ are interaction terms that decay exponentially with the distance between the spins, and $\hat{\tau}^{z}$ are {\it quasilocal} operators that mutually commute~\cite{supmat}. The $\ell$-bit model, \eqref{eq:l-bit},  captures many phenomenological properties of MBL, including eigenvalue statistics~\cite{Sierant20stat, prakash21u}, entanglement~\cite{serbyn2014quantum,Znidarivc18e, aceituno24a}, and other aspects of the dynamics~\cite{Berger24, Szoldra24, Scocco24}. In the following, we utilize \eqref{eq:l-bit} to understand nonstabilizerness dynamics in an MBL system.

We start by analyzing how the SRE grows when different terms are included in~(\ref{eq:l-bit}) for an $X$-polarized initial state $|\Psi_X^{+}\rangle = \bigotimes_{k=1}^{L} (|\downarrow\rangle + |\uparrow\rangle)/\sqrt{2}$. We first focus on the case where the $\ell$-bits do not interact, $J_{ij...}=0$. In this case,~(\ref{eq:l-bit}) describes an Anderson insulator, and the dynamics of the Pauli strings are governed solely by the spin precession. The SRE exhibits rapid initial growth at $t \sim 1$ and saturates fast to a nearly constant value (after averaging over disorder realizations). This dynamics can be accurately captured by decomposing the Pauli strings as the product of individual single-spin observables, yielding~\cite{supmat} 
\begin{equation}\label{Eq.SRE_Anderson1}
\mathcal{M}_2 = -\frac{L}{W} \int_{0}^{W}dh\log_2\bigg[1 -\frac{1}{4}\sin^2(4ht)\bigg]
\end{equation}
for the initial $|\Psi_{X}^{+}\rangle$ state. In the limit of $t\rightarrow \infty$, this expression yields $\mathcal{M}_2 \approx L\log_2(8/7)$, shown by (III) red dashed line in Fig.~\ref{Fig:lbit}. For a generic product initial state $|\Psi_R\rangle$, our numerical results show that the SRE rapidly saturates to a constant value not bigger than $\mathcal{M}_2(|\Psi_R\rangle)+L\log_2(8/7)$ (see~\cite{supmat} for an analytical derivation for arbitrary initial states).

\begin{figure*}[t!]
    \centering   \includegraphics[width=0.99\linewidth]{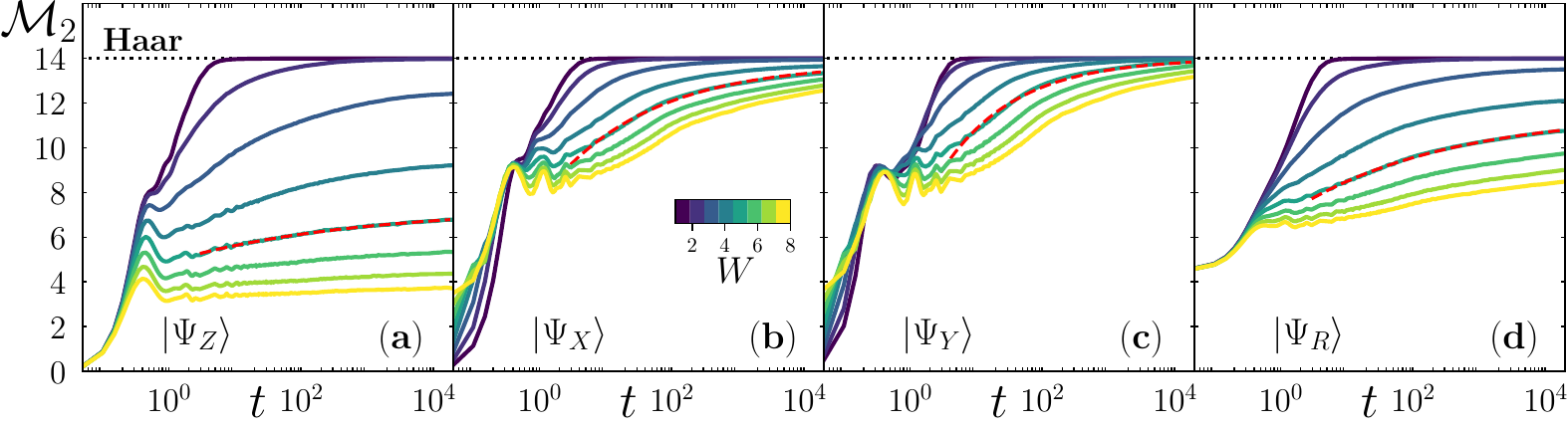}
    \caption{ {\it Dynamics of nonstabilizerness in disordered TFIM.} Evolution of SRE for initial states (a) $|\Psi_Z\rangle$, (b) $|\Psi_X\rangle$, (c) $|\Psi_Y\rangle$, and (d) $|\Psi_R\rangle$ (see text). The results are for $L=16$ and averaged over $1000$ realizations, considering the product state close to the middle of the spectrum. To demonstrate the validity of~\eqref{Eq:SRE_MBL}, we performed a numerical fit at $W=5$ for all states. The saturation value $\mathcal{M}_2^{\mathrm{sat}}$ depends on the initial state: $\mathcal{M}_2^{\mathrm{sat}}\approx7.3$ for $|\Psi_Z\rangle$, $\mathcal{M}_2^{\mathrm{sat}}\approx11.53$  for $|\Psi_R\rangle$, while the SRE for $|\Psi_{X}\rangle$ and $|\Psi_{Y}\rangle$ states saturates to the Haar value $\mathcal{M}_2^{\mathrm{Haar}}$. The power-law growth exponents are $\beta\approx0.16$ for $|\Psi_Z\rangle$, $\beta\approx0.29$ for $|\Psi_{X}\rangle$, $\beta\approx0.39$ for $|\Psi_{Y}\rangle$, and $\beta\approx0.19$ for $|\Psi_{R}\rangle$. Similar behavior is obtained for other disorder strengths within the MBL regime.}
    
    \label{Fig:TFIM}
\end{figure*}

The behavior of SRE significantly changes when the $\ell$-bits interact and $J_{ij...}\neq 0$. Similar to the non-interacting case, the spin precession terms induce a fast growth of the SRE at $t\sim1$. Subsequently, in the presence of interactions, $\mathcal{M}_2$ continues to grow until it reaches a saturation value after sufficiently long times. 

To understand the dynamics of SRE, it is essential to examine the impact of the spin dephasing terms on the time evolution of Pauli strings expectation values~\cite{serbyn2014quantum}. The expectation values of Pauli strings composed solely of $\hat{\tau}^{z}$ operators remain constant under the dynamics of $\hat {\mathcal{H}}_{\ell\mathrm{-bit}}$. On the other hand, Pauli strings containing
$\hat{\tau}^{x}$ or $\hat{\tau}^{y}$  exhibit distinctly different behavior. The slow dephasing of the spins causes a power-law decay of the single-spin expectation values $|\hat{\tau}_j^{\alpha}|$, with $\alpha \in \{x,y\}$; for multi-spin observables, the situation depends on whether the spins corresponding to $\hat{\tau}^{\alpha}$ operators are entangled. If these spins are entangled, the expectation value decays in the same power-law fashion as for the single-spin observables. However, if these spins are not entangled, the expectation value decays as a product of individual single-spin observables and, therefore, decays much faster. 
Before the spins get entangled, the sum of all Pauli strings results in a stretched exponential behavior of the sum in Eq.~\eqref{eq:sre1}, leading to a power-law growth of $\mathcal{M}_2$ with a certain exponent $\beta$. 
As the particles gradually become entangled, this exponent decreases, and the growth of the SRE slows down with time. This results in the following dynamics of SRE in the MBL regime
\begin{equation}\label{Eq:SRE_MBL}    
    \mathcal{M}_2^{\mathrm{MBL}} = \mathcal{M}_2^{\mathrm{sat}} - c/t^{\beta},
\end{equation}
where $\mathcal{M}_2^{\mathrm{sat}}$, $c$ and $\beta$ are constants dependent on the initial state. In particular, for the $X$-polarized initial state $|\Psi_{X}^{+}\rangle$, the saturation value $\mathcal{M}_2^{\mathrm{sat}}= \mathcal{M}_2^{\mathrm{Haar}}$ is the same as for the ergodic dynamics, and $\beta= \beta^{\prime} \ln(2)$ is the fastest exponent for all possible initial configurations. Its dynamics is illustrated by the (I) 
orange dashed line in Fig.~\ref{Fig:lbit} (further discussion can be found in the End Matter section).
Moreover, for a generic initial product state, the saturation value is smaller, $\mathcal{M}_2^{\mathrm{sat}} < \mathcal{M}_2^{\mathrm{Haar}}$ and $\beta < \beta^{\prime} \ln(2)$, as depicted in (II) orange dashed line in Fig.~\ref{Fig:lbit}.

\paragraph*{Microscopic model.} 
To assess the accuracy of the {$\ell$}-bit in capturing the nonstabilizerness dynamics of strongly interacting disordered systems, we analyze a microscopic model expected to exhibit an MBL phase at sufficiently strong disorder, the disordered TFIM, with Hamiltonian 
\begin{equation}
    \hat{\mathcal{H}}_{\mathrm{TFIM}} = \sum_{i=1}^{L-1}  J_{i,i+1} \hat{Z}_i \hat{Z}_{i+1} + \sum_{i=1}^{L} h_i\hat{Z}_i + g \sum_{i=1}^{L}  \hat{X}_i 
    \label{TFIM}
\end{equation}
where $h_i \in [-W,W]$ are random on-site fields that are drawn from a uniform distribution, $g$ is the transverse field, and $J_{i,i+1}$ are the interactions between neighboring spins. Building on~\cite{Imbrie16a}, we consider nearest-neighbor couplings drawn from a uniform distribution $J_{i,i+1} \in [0.8,1.2]$, and fix the transverse field at $g=1$. Mathematical arguments~\cite{Imbrie16a, DeRoeck24absence} suggest that this model hosts an MBL phase for sufficiently large disorder strength $W$. Finite-size scaling analysis~\cite{abanin21d, Tomasi21} places the critical disorder threshold at $W_c \approx 3.5$. However, similar to the XXZ chain~\cite{Sierant20p}, the model exhibits finite-size drifts~\cite{Sierant23st}. We study the quench dynamics of an initial state $|\Psi\rangle$ using the Chebyshev polynomial expansion~\cite{tal84a,Sierant22c} up to $t=2\times10^{4}J$. We consider chains of $L\in[8,20]$ spins, with results averaged over $1000$ disorder realizations. The chosen initial state $|\Psi\gamma\rangle$ is a product state in the $\gamma$ basis ($\gamma \in \{X,Y,Z\}$), with each qubit randomly assigned as $|\pm\rangle_{\gamma}$, and the total energy of  $|\Psi\gamma\rangle$ is close to the middle of the spectrum.

In Fig.~\ref{Fig:TFIM}(a), we show the time evolution of a $Z$ polarized initial state $|\Psi_Z\rangle$. For weak disorder, $W\sim 1$, $\mathcal{M}_2$ quickly grow towards the Haar value in the weak disorder regime, consistent with the ergodic dynamics results~\cite{Turkeshi25spreading, Tirrito24anti}. For strong disorder $W\gtrsim 5$, $|\Psi_Z\rangle$ is close to an eigenstate of~\eqref{TFIM} and, therefore, SRE increases very slowly. In the MBL regime, the behavior of SRE is accurately captured by the phenomenological formula~\eqref{Eq:SRE_MBL} (red dashed line in Fig.~\ref{Fig:TFIM}(a)), with the saturation value $\mathcal{M}_2^{\mathrm{sat}}$ considerably smaller than $\mathcal{M}_2^{\mathrm{Haar}}$.

In Fig.~\ref{Fig:TFIM}(b), we present the SRE evolution of the initial $X$-polarized state $|\Psi_X\rangle$. The SRE, $\mathcal{M}_2$, initially grows rapidly before slowing down at longer times. Even for strong disorder, the SRE approaches the Haar value at late times. A similar trend is observed for $|\Psi_Y\rangle$,  as shown in Fig.~\ref{Fig:TFIM}(c). Additionally, we examine product states constructed by random rotations on the Bloch sphere, denoted as $|\Psi_R\rangle$. Since such states cannot be constructed by Clifford gates, $\mathcal{M}_{2}\neq0$ at $t=0$. However, this initial value is significantly below the Haar limit, as the SRE of product states is limited to $\mathcal{M}_{2}=L\log_2(4/3)$~\cite{Leone2022stabilizer}.
Therefore, under Hamiltonian dynamics, the magic resources spread over time, consistent with \eqref{Eq:SRE_MBL}, and the SRE growth is intermediate between the dynamics for the $|\Psi_Z\rangle$ and $|\Psi_Y\rangle$ states.

\paragraph*{Dependence of the initial state.} To explain the dependence of $\mathcal{M}_2^{\mathrm{sat}}$ on the choice of the initial state, we revisit the $\ell$-bit model. Starting from a random product state in the computational basis $|i^{\prime}\rangle$, we prepare the state
\begin{figure}[t!]
    \centering
    \includegraphics[width=1\linewidth]{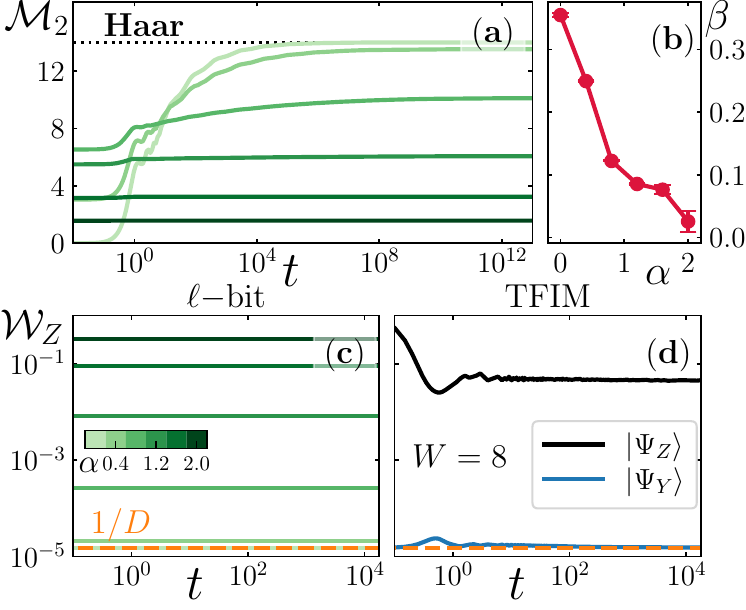}
    \caption{ 
    {\it Initial state dependence of SRE and weight of Z gates in MBL regime.} (a) SRE dynamics in the $\ell$-bit model for different choices of $\alpha$. (b) Power-law exponent $\beta$ characterizing the growth of $\mathcal{M}_2$ as a function of $\alpha$; time evolution of $\mathcal{W}_Z$ for different initial states for the (c) $\ell$-bit model and (d) TFIM. The saturation value of $\mathcal{M}_2$ depends on the degree of localization of the initial state in the $\ell$-bit basis and it is intrinsically connected to $\mathcal{W}_Z$.}    
    \label{Fig:WZ_main}
\end{figure}
\begin{equation}
    |\Psi_{H}\rangle = \frac{1}{\sqrt{Z}}\sum_{i=1}^{D} e^{-\alpha {d}(i,i^{\prime})}|i\rangle, \quad Z = \sum_{i=1}^{D} e^{-2\alpha d(i,i^{\prime})}
    \label{Hamming_State}
\end{equation}
where $d(i,i^{\prime})$ is the Hamming distance between two states and $\alpha$ is a parameter that controls the degree of localization in the computational basis. In Fig.~\ref{Fig:WZ_main}(a), we show that the dynamics of $\mathcal{M}_2$ slows down as $\alpha$ increases, i.e. when the initial state becomes more localized in the computational basis. Saturation values $\mathcal{M}_2^{\mathrm{sat}}$ also critically depend on the value of $\alpha$. The power-law exponent $\beta$, as illustrated in Fig.~\ref{Fig:WZ_main}(b), decreases monotonically as $\alpha$ increases, practically vanishing for $\alpha\geq2$.

The reason for different values of $\mathcal{M}_2^{\mathrm{sat}}$ becomes clear if we write $\mathcal{M}_2 =-\log_2(\mathcal{W}_Z + c)$, where $\mathcal{W}_Z$ is the contribution from the Pauli subgroup $\mathcal{P}_{IZ}$ containing only $\hat{I}$ and $\hat{Z}$ gates, whereas $c$ takes into account the remaining strings. It is easy to show (see the End Matter) that $\mathcal{W}_Z$ can be written as 

\begin{equation}
    \mathcal{W}_{Z}(t) = \sum_{u,v,k \in \{0,1\}^L} |c_u(t)|^2|c_v(t)|^2|c_k(t)|^2|c_{u\oplus v\oplus k}(t)|^2,
    \label{W_Z}
\end{equation}

As mentioned before, Pauli strings from $\mathcal{P}_{IZ}$ remain frozen during the dynamics of~\eqref{eq:l-bit}, so $\mathcal{W}_Z$ is a constant fixed by the initial state, as shown in Fig.~\ref{Fig:WZ_main}(c). Similar behavior is obtained for the TFIM deep in the MBL regime ($W=8$), where $\tau_{i}^{z}\approx Z_i$. As illustrated in Fig.~\ref{Fig:WZ_main}(d), the choice of $|\Psi_Y\rangle$ in the TFIM leads to $W_Z\approx1/D$, in agreement with the completely delocalized case in the $\ell$-bit. On the other hand, the choice of $|\Psi_Z\rangle$ leads to very slow dynamics of $\mathcal{W}_Z$, showing that this state is close to an eigenstate of~\eqref{eq:l-bit} (see the End Matter for further discussion).

\paragraph*{Collapsing of stabilizer entropy.}

\begin{figure}[t!]
    \centering
    \includegraphics[width=0.99\linewidth]{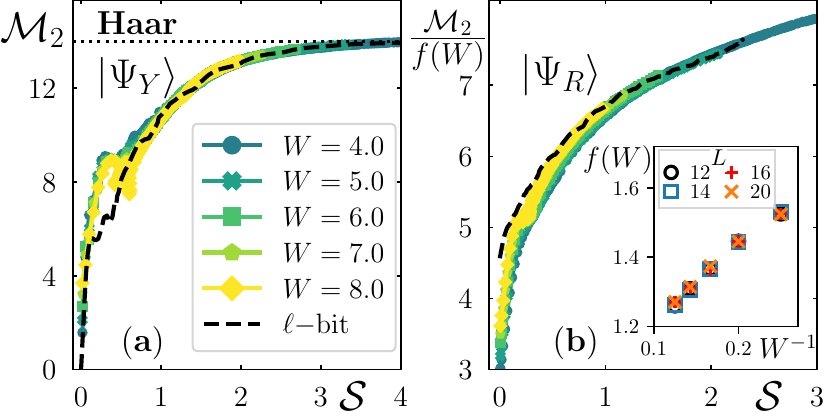}
    \caption{ {\it Nonstabilizerness versus entanglement in the MBL regime.} The SRE $\mathcal{M}_2$ is plotted as a function of the half-chain entanglement entropy $\mathcal{S}$ for different disorder strengths $W$ and system size $L$. For the $Y$-polarized state $|\Psi_Y\rangle$, (\textbf{a}), $\mathcal{M}_2(\mathcal{S})$ collapse, without any fitting parameters, on a single master curve both for $\ell$-bit model and TFIM. For the random product state $|\Psi_R\rangle$, (\textbf{b}), the collapse occurs when $\mathcal{M}_2(\mathcal{S})$ are rescaled by an $L$-independent factor $f(W)$.
    }
    \label{Fig:Magic_Ent}
\end{figure}

To understand the interplay of nonstabilizerness and entanglement in the MBL regime, we analyze in Fig.~\ref{Fig:Magic_Ent} the growth of $\mathcal{M}_2$ as a function of the half-chain entanglement entropy $\mathcal{S}=-\mathrm{tr}(\rho_{L/2}\ln{\rho_{L/2}})$, where $\rho_{L/2}$ is the reduced density matrix obtained by tracing out the degrees of freedom of the first half of the system. The entanglement entropy has been proposed as an ``internal clock'' for disordered localized systems, providing a natural way to compare the dynamic evolution at different values of $W$~\cite{Evers23Internal}. 

We consider $|\Psi_Y\rangle$ as our initial state. As shown in Fig.~\ref{Fig:Magic_Ent}~(a), both $\ell$-bit model and TFIM exhibit similar dynamics, with $\mathcal{M}_2(\mathcal{S})$ collapsing onto a single master curve without any scaling parameters. For an initial random product state $|\Psi_{R}\rangle$, the $\ell$-bit model predictions are aligned with the TFIM results when a disorder-dependent rescaling function, $f(W)$,  is introduced. In Fig.~\ref{Fig:Magic_Ent}(b) we show $\mathcal{M}_2 / f(W)$ for $|\Psi_{R}\rangle$, where $f(W)$ is found by minimizing deviations of TFIM results from the $\ell$-bit model predictions. Importantly, $f(W)$ is independent of system size $L$. 
These results demonstrate a close connection between the SRE and the growth of entanglement in the MBL regime, revealing that the correlations built within the MBL are genuinely quantum, arising from non-Clifford operations. 
The link between entanglement and nonstabilizerness observed for MBL parallels results in ergodic systems~\cite{odavic24s,Tirrito24anti,Hou25highway}, but it is not generic; e.g.,  scar eigenstates of the PXP model~\cite{smith2024non} exhibit significant SRE despite limited entanglement (see the End Matter for a more detailed discussion).

\paragraph*{Conclusions and outlook.}
In this work, we have explored the dynamics of nonstabilizerness in disordered many-body systems exhibiting MBL.
By developing a theoretical framework founded on an $\ell$-bit phenomenology, we have shown that nonstabilizerness spreading in MBL systems is fundamentally constrained by the slow dynamics characteristic of the MBL regime. Unlike ergodic systems, where SRE rapidly saturates to its maximal value, MBL systems display a much slower power-law relaxation towards the saturation value, which exhibits strong variability with respect to the choice of the initial state. Through numerical simulations of the disordered TFIM, we have verified our phenomenological formulas for the SRE growth and demonstrated a strong connection between the SRE and entanglement entropy growth in the MBL regime. In particular, our findings show that the disorder suppresses the rapid spread of magic resources, and interactions play a crucial role in enabling its slow but sustained growth.

Our results open up several intriguing directions for future research. One promising avenue is investigating the interplay between nonstabilizerness and other forms of ergodicity breaking, such as quantum many-body scars~\cite{serbyn2021quantum,smith2024non,hartse2024stabilizer} and disorder-free localization~\cite{PhysRevLett.118.266601, Brenes18,vanNieuwenburg19,Schulz19,Yao20,Yao21,Yao21t}. In particular, extending our analysis to gauge theories~\cite{Banuls2020,Aidelsburger21} and constrained quantum systems~\cite{Chen18, Sierant21constr, Theveniaut20, Pietracaprina21, Royen24} could shed further light on the role of nonstabilizerness in non-thermalizing quantum dynamics.
Moreover, this also motivates further explorations of Clifford-augmented matrix product states~\cite{masotllima2024stabilizertensornetworks, Qian24prl, lami2024quantumstatedesignsclifford,qian2024cliffordcircuitsaugmentedtimedependent, nakhl2024sta} in scenarios where continuous Hamiltonians govern the dynamics. Such investigations may provide a broader understanding of the interplay of disorder and nonstabilizerness spreading for quantum error correction and fault-tolerant quantum computing.

\begin{acknowledgments}
\paragraph*{Acknowledgments.}
E.T. and P.S. acknowledge collaboration with X. Turkeshi and M. Dalmonte on related subjects. P.S. acknowledges insightful discussions with P. Stornati, S. Masot-Llima, and  A. Garcia-Saez.
The work of P.R.N.F. and J.Z. was partially funded by the National Science Centre, Poland, project 2021/03/Y/ST2/00186 within the QuantERA II Programme (DYNAMITE) that has received funding from the European Union Horizon 2020 research and innovation programme under Grant agreement No 101017733. The work of P.R.N.F. and J.Z. was also funded by the National Science Centre, Poland, project  2021/43/I/ST3/01142 -- OPUS call within the WEAVE programme.  
P.S. acknowledges fellowship within the “Generación D” initiative, Red.es, Ministerio para la Transformación Digital y de la Función Pública, for talent attraction (C005/24-ED CV1), funded by the European Union NextGenerationEU funds, through PRTR.
E.\,T. acknowledges support from  ERC under grant agreement n.101053159 (RAVE), and
CINECA (Consorzio Interuniversitario per il Calcolo Automatico) award, under the ISCRA
initiative and Leonardo early access program, for the availability of high-performance computing resources and support.
The study was also partially funded by the ``Research Support Module'' as part of the ``Excellence Initiative – Research University'' program at the Jagiellonian University in Kraków.
We gratefully acknowledge the Polish high-performance computing infrastructure PLGrid (HPC Centers: ACK Cyfronet AGH) for providing computer facilities and support within the computational grant no. PLG/2024/017289
\end{acknowledgments}

\paragraph{Data Availability.}- The data that support the findings of this article are openly available~\cite{UJ/WMN9VI_2025}.

\normalem

\newpage

\onecolumngrid 
\section*{End Matter} 
\twocolumngrid 

In this work, we explored the dynamics of nonstabilizerness, or ``magic'', in many-body localized (MBL) systems. Using a combination of analytical and numerical methods, we derived and verified power-law saturation of stabilizer Rényi entropy (SRE) evolution in the MBL regime. Our findings highlight the critical role of interactions and disorder in modulating the growth and saturation of nonstabilizerness, offering clear distinctions between ergodic, Anderson localized, and MBL regimes.

\begin{figure}[b!]
    \centering
    \includegraphics[width=1\columnwidth]{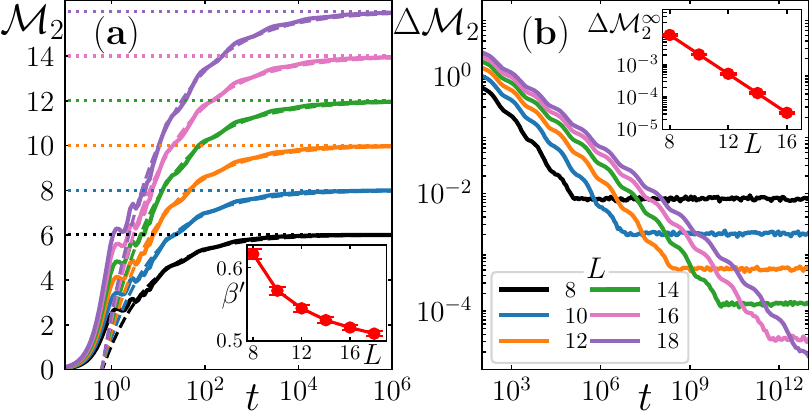}
    \caption{(a) Nonstabilizerness spread in the $\ell$-bit model for different system sizes, with $|\Psi_{X}^{+}\rangle$ as the initial state. The dashed lines show the analytical solution (Eq.~\eqref{Eq:SRE_MBL} of the main text), which accurately describes the SRE growth after $t\sim 1$. Inset: Dependence of the power-law exponent $\beta^{\prime}$ with the system size for $\xi=0.5$; (b) Time evolution of $\Delta \mathcal{M}_2$ for different system sizes $L$. It decays polynomially in $t$ until it eventually saturates, except for $L=18$, for which the Heisenberg time is beyond the considered here. The saturation value exhibits an exponential decay with exponent $\lambda \approx -\ln{2}$, as shown in the inset. }
    \label{Fig:SizeEffects1}
\end{figure}

\paragraph{Accuracy of the Power-law Description of the SRE.}
In this section, we validate the analytical power-law behavior given by Eq.~\eqref{Eq:SRE_MBL}. Specifically, we demonstrate its accuracy by comparing analytical predictions to numerical simulations of SRE dynamics in MBL systems.
In order to evaluate the accuracy of~\eqref{Eq:SRE_MBL}, we assume $|\Psi_{X}^{+}\rangle$ as the initial state and compute the SRE dynamics for various system sizes $L$, keeping the localization length $\xi=0.5$ (see~\cite{supmat} for more details on the $\ell$-bit Hamiltonian). As illustrated in Fig.~\ref{Fig:SizeEffects1}(a), the SRE exhibits a power-law dependence on time at time-scales where entanglement is relevant, saturating at a size-dependent value (dotted lines). The ``bump'' observed at early times ($t\sim 1$) is attributed to the initial spin precession that is reminiscent of the non-interacting dynamics. As illustrated in the inset plot of Fig.~\ref{Fig:SizeEffects1}(a), the power-law exponent $\beta^{\prime}$ has a non-trivial system size dependence. This exponent is directly related to the localization length $\xi$ of~\eqref{eq:l-bit}, although its exact relationship remains an outstanding open problem~\cite{serbyn2014quantum, Abanin19}.

To further analyze the saturation behavior of $\mathcal{M}_2$, we introduce the deviation from the Haar value, defined as $\Delta\mathcal{M}_2 = \mathcal{M}_{2}^{\mathrm{Haar}} - \mathcal{M}_{2}$. 
Fig.~\ref{Fig:SizeEffects1}(b) illustrates that this saturation value decays exponentially with system size, characterized by the exponent $\lambda \approx -\ln{2}$.

This finite-size dependence observed in the saturation value of the SRE can be attributed primarily to the dynamics of $\mathcal{W}_Z$. At sufficiently long times, the spins become completely dephased, making the expectation values of strings involving $X$ and $Y$ gates indistinguishable from those of a random state. However, in finite-size systems, the weight $\mathcal{W}_Z$—associated exclusively with identity and $Z$ operators—differs notably. Despite this finite-size discrepancy, we conjecture that both contributions scale as $1/D$ and thus will converge identically in the thermodynamic limit.

\paragraph{Derivation and Dynamics of $\mathcal{W}_Z$.}

In the main text, we showed that the initial state determines the saturation value of the SRE and, in the strong disorder limit, $\mathcal{M}_2$ can be analytically derived. Here, we detail the analytical derivation of the weight of $Z$ gates, $\mathcal{W}_Z$, and examine its dynamical behavior in the ergodic and MBL regimes of the TFIM.

The weight of $Z$ gates is defined as:
\begin{equation}\label{Eq:WZ}
    \mathcal{W}_{Z}(|\Psi\rangle) = \sum_{P\in\mathcal{P}_{IZ}} \frac{|\langle \Psi|  P |\Psi\rangle|^{4}}{D}
\end{equation}
where $\mathcal{P}_{IZ}$ is the subgroup of the Pauli group $\mathcal{P}_L$ consisting exclusively of identity $\hat{I}$ and $\hat{Z}$ operators. Any Pauli operator in this subgroup can be expressed as $P_w= \otimes_{k=1}^{L} Z_{k}^{w_k}$ with $w\in\{0,1\}^{L}$. For a time-evolved state $|\Psi(t)\rangle = \sum_{u\in\{0,1\}^L} c_{u}(t)|u\rangle$, the expectation value of $P_w$ is

\begin{equation}
    \langle \Psi|P_w|\Psi\rangle = \sum_{u} |c_u(t)|^2(-1)^{w\cdot u}|u\rangle
    \label{Eq:Expect_Z}
\end{equation}

Substituting Eq.~\eqref{Eq:Expect_Z} into Eq.~\eqref{Eq:WZ}, the contribution involving solely $I$ and $Z$ gates is
\begin{equation}
    \mathcal{W}_Z = \frac{1}{2^L}\sum_{u,v,k,l}|c_u(t)|^2|c_v(t)|^2|c_k(t)|^2|c_l(t)|^2\sum_{w}(-1)^{w\cdot s}
    \label{Eq:WZ2}
\end{equation}
with $s=u+v+k+l$. The summation over $w$ simplifies to

\begin{equation}
\sum_{w\in\{0,1\}^{L}} (-1)^{w\cdot s}
        = 2^{L}\,\delta_{s\bmod 2,0},
\end{equation}

leading to the constraint $l=u\oplus v\oplus k$. Therefore, $\mathcal{W}_Z$ is given by 

\begin{equation}
    \mathcal{W}_{Z}(t) = \sum_{u,v,k \in \{0,1\}^L} |c_u(t)|^2|c_v(t)|^2|c_k(t)|^2|c_{u\oplus v\oplus k}(t)|^2,
    \label{W_Z}
\end{equation}

\begin{figure}[t!]
    \centering
    \includegraphics[width=0.85\linewidth]{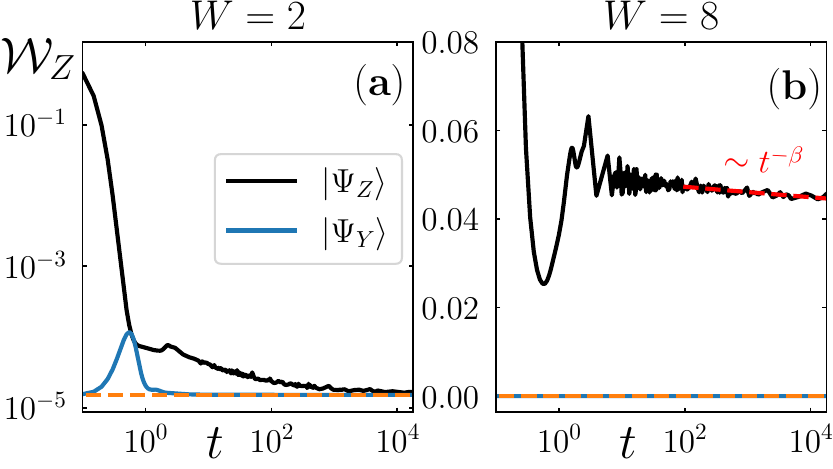}
    \caption{$\mathcal{W}_Z$ dynamics for two different states, $|\Psi_Y\rangle$ and $|\Psi_Z\rangle$, in the (a) ergodic and (b) localized regimes of the TFIM model. In the simulation, we consider $L=16$ and average the results over, at least, 1000 realizations. In the ergodic regime, the two different states evolve similarly due to the lack of integrals of motion. In the localized regime, however, the weight of $Z$ gates ($\mathcal{W}_Z$) depend on the initial state, leading to different saturation values of the SRE.}
    \label{Fig:Wzs}
\end{figure}

Since the MBL regime is characterized by an extensive set of integrals of motion, any string from the subgroup $\mathcal{P}_{IZ}$ remains constant throughout the entire evolution. In order to confirm our microscopic model aligns with the $\ell$-bit phenomenology, we must contrast the dynamics of $\mathcal{W}_Z$ in both regimes of the TFIM.
In Fig.~\ref{Fig:Wzs}(a)), we show that $\mathcal{W}_Z$ quickly approaches the value $1/D$ for a random product state in the $Z$ basis, $|\Psi_Z\rangle$. For an initial state polarized along the $Y$ basis, $|\Psi_Y\rangle$, $\mathcal{W}_Z$ remains nearly constant, indicating a delocalized behavior.
In contrast, within the MBL regime (Fig.~\ref{Fig:Wzs}(b)), the dynamics significantly differ: for $|\Psi_Y\rangle$, $\mathcal{W}_Z$ stays near the ergodic value, while for $|\Psi_Z\rangle$, it remains almost constant, consistent with the $\ell$-bit phenomenology. Although data suggest a small power-law decay with exponent $\lambda=0.011(2)$ at intermediate times, indicative of potential slow thermalization, extending the analysis to later times shows a diminishing exponent, hinting at eventual saturation at the Heisenberg time.

\begin{figure}[t!]
    \centering
    \includegraphics[width=0.99\linewidth]{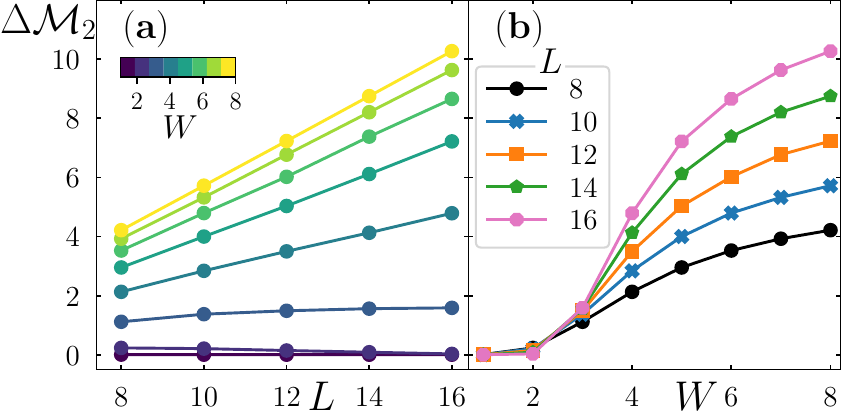}
    \caption{ {\it Nonstabilizerness across the ETH-MBL crossover in TFIM.} (a) Deviation from the Haar value $\Delta\mathcal{M}_2$ as a function of system size $L$ and for different disorder strengths. (b) $\Delta\mathcal{M}_2$ as a function of disorder strength  $W$ at the latest time available for the $|\Psi_{Z}\rangle$ distinguishes the ETH and MBL regimes. }
    \label{Fig:TFIM2}
\end{figure}

\paragraph{ETH-MBL crossover-} We now investigate whether $\mathcal{M}_2$ can serve as a tool for distinguishing the crossover between ETH and MBL regimes. The SRE has been linked to phase transitions in many ground-state problems~\cite{tarabunga23m,tarabunga2023magic,falcao24a,tarabunga2024critical,ding2025evaluating} and, therefore, one may expect that it is sensitive to the ETH-MBL crossover observed for $L\approx 20$ at $W_c\sim 3.5$~\cite{abanin21d}. We consider the time evolution of $|\Psi_Z\rangle$ and compute $\Delta\mathcal{M}_2 = \mathcal{M}_2^{\mathrm{Haar}} - \mathcal{M}_2$ for different size systems $L$ at the longest available time ($t=2\times 10^{4}$), as shown in Fig.~\ref{Fig:TFIM2}(a). For weak disorder strength,  $\Delta\mathcal{M}_2 \rightarrow 0$ as $L$ increases, as expected for the ETH regime. On the other hand, $\Delta\mathcal{M}_2$ grows linearly with $L$ for strong disorder. The crossover between these two regimes occurs near $W_c\approx 3.5$, showing how the SRE distinguishes the ergodic and non-ergodic dynamical regimes. 

\paragraph{Relation between entanglement entropy and nonstabilizerness-}
One of the central result of this work is the observation that, in the MBL regime, the dynamics of the SRE and the entanglement entropy collapse onto a single universal curve, independent of disorder strength and system size (see Fig.~\ref{Fig:Magic_Ent}). 

This strong correlation is highly nontrivial, as the two quantities capture distinct aspects of quantum correlations: entanglement entropy quantifies the amount of bipartite entanglement, while SRE measures the non-Clifford complexity—the genuinely quantum, classically intractable part—of those correlations. Their common scaling thus implies that the slow entanglement growth in MBL systems is accompanied by the gradual accumulation of nonstabilizerness.

Our findings provide a resource-theoretic lens on localization in interacting systems: disorder constrains not only transport and thermalization but also the generation of magic resources. This connection is not universal across all non-ergodic systems. For instance, scarred eigenstates in the PXP model display significant SRE despite limited entanglement~\cite{smith2024non}. The link between entanglement and nonstabilizerness observed here parallels findings in ergodic systems~\cite{odavic24s,Tirrito24anti,Hou25highway}, while its emergence in a non-ergodic regime highlights the genuinely quantum complexity of MBL dynamics.

\newpage

\newcommand{\snum}{S}

\renewcommand{\theequation}{\snum.\arabic{equation}}
\renewcommand{\thefigure}{\snum.\arabic{figure}}

\setcounter{equation}{0}
\setcounter{figure}{0}

\newpage 
\section*{Supplementary material: Nonstabilizerness dynamics in many-body localized systems}
\label{appendix1}

In this section, we demonstrate how to obtain an analytical expression for the stabilizer R\'enyi entropy (SRE) for a perfect Anderson insulator. In this case, the $\ell$-bits do not interact with each other and, therefore, only the first term of $\hat{\mathcal{H}}_{\mathrm{\ell-bit}}$ remains relevant. The SRE can be rewritten as
\begin{equation}\label{Eq.Avg_SRE}
    \mathcal{M}_2 = \mathbb{E}_{\{h\}} \Bigg(-\log_2 \Bigg[\sum_{P \in \mathcal{P}_L} \frac{|\mathrm{Tr}(\rho P)|^{4}}{D} \Bigg]\Bigg)
\end{equation}
where $\mathbb{E}_{\{h\}}$ denotes the average over different disorder realizations and $\rho$ is the density matrix of the state $|\Psi(t)\rangle$. Since the system is noninteracting, the density matrix can be written as $\rho = \bigotimes_{k=1}^{L}\rho_k$, where $\rho_k$ is the reduced density matrix of the $k$-th spin. For any initial state $|\Psi\rangle = \bigotimes_{k=1}^{L}(\cos(\theta_k/2)|\uparrow\rangle + e^{i\phi_k}\sin(\theta_k/2)|\downarrow\rangle)$, the reduced density matrix $\rho_k$ at time $t$ is given by~\cite{Serbyn14}:
\begin{equation}\label{Eq.Density_k}
    \rho_k(t) = \begin{pmatrix}
    \cos^2(\theta_k/2) & \frac{\sin\theta_k}{2}e^{-i(2h_kt + \phi_k)} \\
    \frac{\sin\theta_k}{2}e^{i(2h_kt +\phi_k)} & \sin^2(\theta_k/2)
    \end{pmatrix}
\end{equation}
where $h_k$ is the local magnetic field at site $k$. In this case, the sum of expectation values of all Pauli strings can be rewritten as 
\begin{equation}\label{Single}
    \sum_{P\in\mathcal{P}_L}|\mathrm{Tr}(\rho P)|^4 = \prod_{k=1}^{L}\sum_{\sigma\in\{I,X,Y,Z\}} |\mathrm{Tr}(\rho_k\sigma_k)|^4
\end{equation}
This implies that, to understand how the SRE grows in the noninteracting picture, it suffices to understand the dynamics of single-spin observables. The single qubit SRE can be easily obtained through the reduced density matrix $\rho_k$, yielding

\begin{equation}
    \sum_{{\sigma\in\{I,X,Y,Z\}}} |\mathrm{Tr}(\rho_k\sigma_k)|^4 = 
    \frac{1}{2}(4 - \sin^2(2\theta_k)
    - \sin^4\theta_k\sin^2\delta_k)
\end{equation}
where $\delta_k = 4h_kt + 2\phi_k$. Substituting this expression into Eq.~\ref{Single}, and assuming that the on-site fields $h_k$ are drawn independently and identically from a uniform distribution in the interval $[-W,W]$, we obtain 

\begin{multline}\label{Eq.SRE_Single}
\mathcal{M}_2 = -\sum_{k=1}^{L}\frac{1}{W}\int_{0}^{W} dh\,\log_2\biggl[\,1 - \frac{1}{4}\sin^2(2\theta_k) \\ 
- \frac{1}{4}\sin^4\theta_k\sin^2\delta_k\biggr].
\end{multline}
where the first term inside the bracket comes from the identity contribution, the second to the expectation value of $\tau^{z}$ operators, and the third term is obtained from the $\tau^{x/y}$ operators. At $\theta=\pi/2$, the second term vanishes and the single-qubit SRE significantly differs from its initial value. To quantify this change, we define the nonstabilizerness gain:

\begin{figure}[t!]
    \centering
    \includegraphics[width=1\columnwidth]{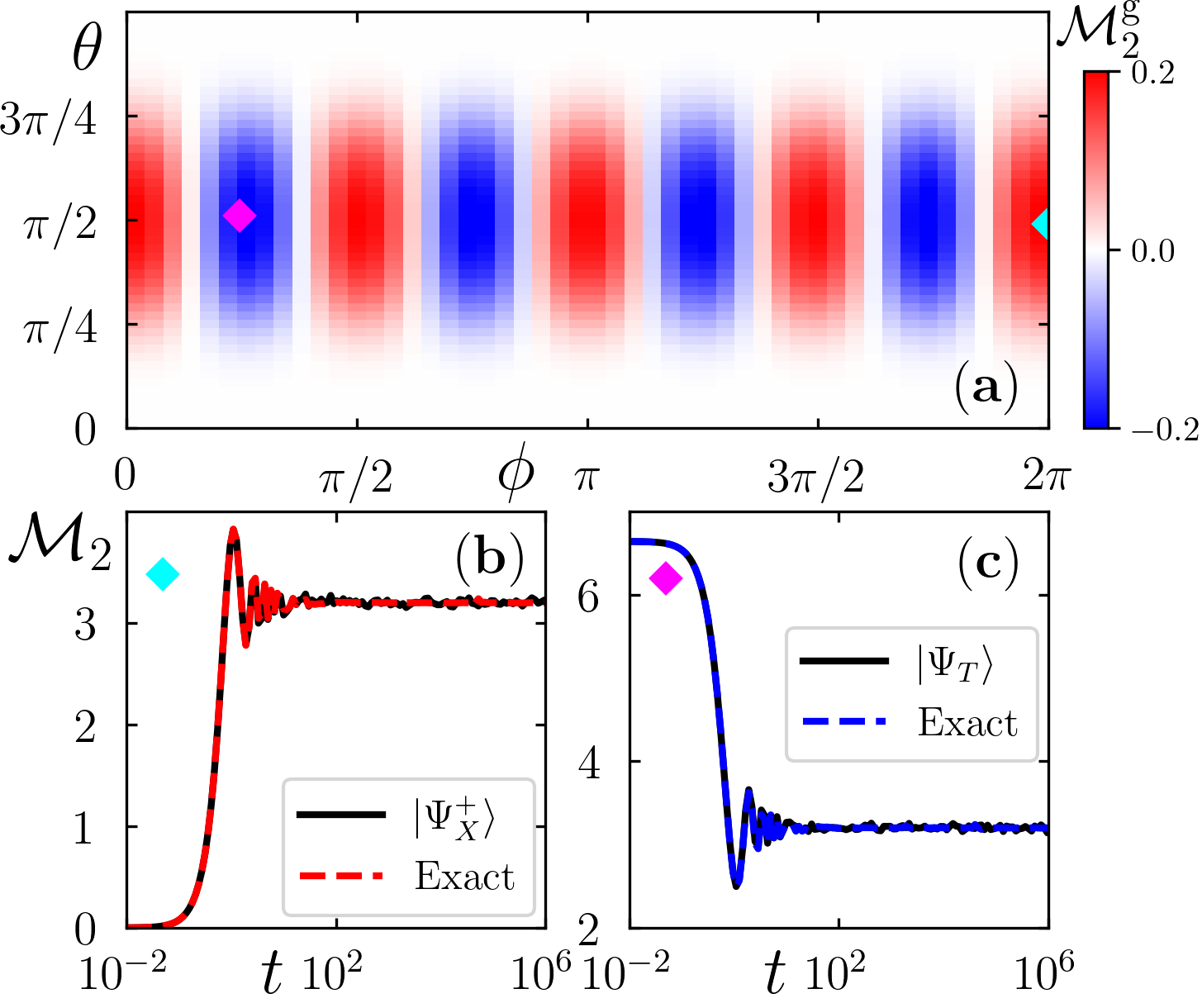}
    \caption{(a) Nonstabilizerness gain (see text) for a single qubit SRE throughout the whole parameter space ($\theta,\phi$). The greatest gain occur for an initial $|+\rangle$ state, while the smallest occur for an initial $|T\rangle$ state. Comparison between the analytical solution given by Eq.~\ref{Eq.SRE_Single} (dashed line) and the numerical calculations of the $\ell$-bit model (solid line) for an initial (b) $|\Psi_X^{+}\rangle$ and (c) $|\Psi_T\rangle$ state. The results were obtained for $L=16$ and $10^{3}$ disorder realizations.  }
    \label{Fig:Anderson}
\end{figure}

\begin{equation}
    \mathcal{M}_2^{\mathrm{g}} = \mathcal{M}_2^{\mathrm{sat}} -\mathcal{M}_2(|\Psi\rangle)
\end{equation}
where $\mathcal{M}_2^{\mathrm{sat}}$ is the asymptotic value of the SRE and $\mathcal{M}_2(|\Psi\rangle)$ its value at $t=0$. In Fig~\ref{Fig:Anderson}(a), we show $\mathcal{M}_2^{\mathrm{g}}$ throughout the parameter space $(\theta,\phi)$, confirming that the highest variance of the SRE is around $\theta=\pi/2$.  

In particular, the $X$ polarized state yields the greatest gain. To confirm our analytical result, we consider $|\Psi_{X}^{+}\rangle = \bigotimes_{k=1}^{L}|+\rangle$ as the initial state, where $|+\rangle$ is an eigenstate of the $\hat{X}$ operator. In this particular case, Eq.~\ref{Eq.SRE_Single} can be written as

\begin{equation}\label{Eq.SRE_Anderson}
\mathcal{M}_2 = -\frac{L}{W} \int_{0}^{W}dh\log_2\bigg[1 -\frac{1}{4}\sin^2(4ht)\bigg]
\end{equation}
which, in the limit $t\rightarrow \infty$, yields $\mathcal{M}_2 \approx L\log_2(8/7)$. In Fig.~\ref{Fig:Anderson}(b), we show a comparison between the analytical solution (dashed red line) and the numerical result (solid black line). The analytical solution perfectly describes the behavior of $\mathcal{M}_2$, showing the accuracy of our analytical calculation. In a similar spirit, we compare the exact and numerical results for an initial state $|\Psi_T\rangle=\bigotimes_{k=1}^{L}|T\rangle$, where $|T\rangle=(1/\sqrt{2})(|0\rangle + e^{i\pi/4}|1\rangle)$ is the magic state. In the disorderless case, $|\Psi_T\rangle$ has the highest SRE of an unentangled state~\cite{Leone2022stabilizer}. Here, however, the initial spin precession quickly decreases its SRE value, leading to a saturation value that significantly differs from its initial value, as shown in Fig.~\ref{Fig:Anderson}(c).

\section{Details on the $\ell$-bit model}

\begin{figure}[t!]
    \centering
    \includegraphics[width=1\columnwidth]{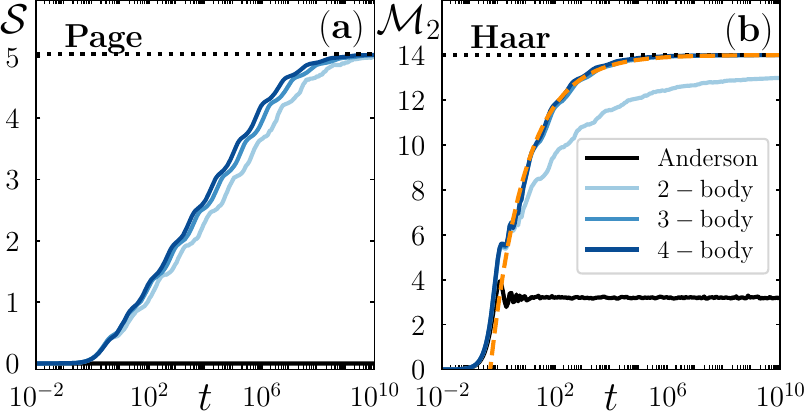}
    \caption{    
    (a) Half chain entanglement entropy ($\mathcal{S}$) and (b) SRE dynamics in the $\ell$-bit model for a chain with $L=16$ spins and localization length $\xi=0.5$. The simulation was performed assuming an initial state $|\Psi(0)\rangle = |+\rangle^{\otimes L}$ and the results were averaged over several disorder realizations.  }
    \label{Fig:EntLBIT}
\end{figure}

In the MBL regime, when the disorder is coupled to on-site operators~\footnote{ If the disorder is not coupled to on-site operators, the standard $\ell$-bit description may fail to capture the system's physics. In such cases, a more sophisticated RSRG-X approach is required to correctly describe the model's properties~\cite{Aramthotill24}}, the system is simply described by the emergence of an extensive large set of integrals of motion, also known as localized bits ($\ell$-bits). 
The {\it quasilocal} structure of the $\ell$-bits encapsulates key signatures of MBL, making the Hamiltonian $\hat{\mathcal{H}}_{\mathrm{\ell-bit}}$ (see Eq. 3 of the main letter) a natural framework to probe the nonstabilizerness dynamics in MBL. 
The physical degrees of freedom are related to the $\ell$-bits by quasilocal unitary transformations $U$, satisfying $\hat{\tau}_{j}^{z} = U\hat{Z}_j U^{\dagger}$. Ideally, $U$ reflects the system's localized nature, with its matrix elements decaying exponentially with the distance between $\ell$-bits. However, for practical purposes, a good approximation is to take $U=I$, where $I$ is the identity operator, significantly simplifying the analysis while retaining the essential features of MBL dynamics.

In our simulations, we model interactions between $\ell$-bits as random variables drawn from a Gaussian distribution with zero mean, where the variance decays exponentially with the distance between spins, following the approach in~\cite{Znidarivc18e}. Specifically, the variance of the two-body interaction term is given by
\begin{equation}
    \langle (J_{i,j})^2 \rangle = e^{-2|j-i|/\xi}
\end{equation}
where $\xi$ is the localization length of the system. A similar exponential decay is assumed for higher-order interactions, considering the maximum separation between spins.

To investigate how different interaction terms in the $\ell$-bit Hamiltonian influence key observables, we first analyze the dynamics of the half-chain entanglement entropy, defined as
\begin{equation}
    \mathcal{S} =  -\mathrm{tr}(\rho_{L/2}\ln{\rho_{L/2}})
\end{equation}
where $\rho_{L/2}$ is the reduced density matrix obtained by tracing out the degrees of freedom of the first half of spins. 
Fig.~\ref{Fig:EntLBIT}(a) shows the growth of $\mathcal{S}$ starting from the initial state $|\Psi_X^{+}\rangle$. In the Anderson insulating case (solid black line), where interactions are absent, correlations cannot spread, and therefore $\mathcal{S}=0$. However, in interacting cases, $\mathcal{S}$ grows logarithmically in time, a hallmark of MBL. The rate of entanglement growth increases slightly as additional interaction terms are included in $\hat{\mathcal{H}}_{\mathrm{\ell-bit}}$ as illustrated in Fig.~\ref{Fig:EntLBIT}(a). The differences between the interaction terms are more striking in the SRE behavior, as shown in Fig.~\ref{Fig:EntLBIT}(b). Assuming only two-body interactions between the $\ell$-bits does not lead to the SRE reaching the Haar value, at least for the time scales analyzed here. However, as we add corrections to the $\ell$-bit Hamiltonian, the dynamics of the SRE approximates the phenomenological prediction described in Eq.(5) of the main letter, as illustrated by the orange dashed line.

\begin{figure}[t!]
    \centering
    \includegraphics[width=1.\columnwidth]{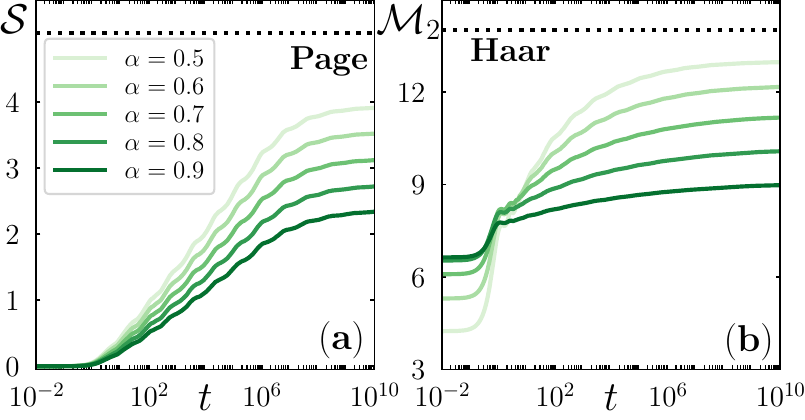}
    \caption{(a) Entanglement entropy and (b) SRE dynamics for a slightly perturbed eigenstate of the system. We consider $L=16$, $\xi=0.5$, and averaged the results over $500$ disorder realizations. Despite the initial value, the dynamics of each state resembles the dynamics of the $|\Psi_Z\rangle$ in microscopic models. }
    \label{Fig:Hamming}
\end{figure}

We also verify how the SRE will spread for an initial state that is close to an eigenstate of the system. To this end, we prepare a random product state in the computational basis and apply a unitary operator that is exponentially localized in the Fock space. Concretely, we select a random product state $|i^{\prime}\rangle$ among $D=2^L$ basis states and construct the state
\begin{equation}
    |\Psi\rangle = \frac{1}{\sqrt{Z}}\sum_{i=1}^{D} e^{-\alpha {d}(i,i^{\prime})}|i\rangle, \quad Z = \sum_{i=1}^{D} e^{-2\alpha d(i,i^{\prime})}
\end{equation}
where $d(i,i^{\prime})$ denotes the Hamming distance between two different strings of the same length. The parameter $\alpha$ controls the degree of localization in the computational basis and, therefore, the dynamics of this state in the $\ell$-bit model should resemble the quench dynamics of an initial $|\Psi_Z\rangle$ in microscopic models, with $\alpha$ playing the role of an effective disorder strength.

In Fig.~\ref{Fig:Hamming}(a), we present the time evolution of the entanglement entropy for different localization strengths $\alpha$. Although $\mathcal{S}$ exhibits the expected logarithmic growth over time, its saturation value remains well below the Page value (black dotted line), indicating that the spins are not fully dephased. The extent of this saturation depends on $\alpha$. A similar dependence on $\alpha$ is observed in the saturation value of $\mathcal{M}_2$. However, as shown in Fig.~\ref{Fig:Hamming}(b), the overall time-scaling behavior of $\mathcal{M}_2$ remains unchanged, suggesting that while localization strength influences the final nonstabilizerness content, the fundamental growth dynamics are robust to variations in  $\alpha$.

\begin{figure}[h!]
    \centering
    \includegraphics[width=1\linewidth]{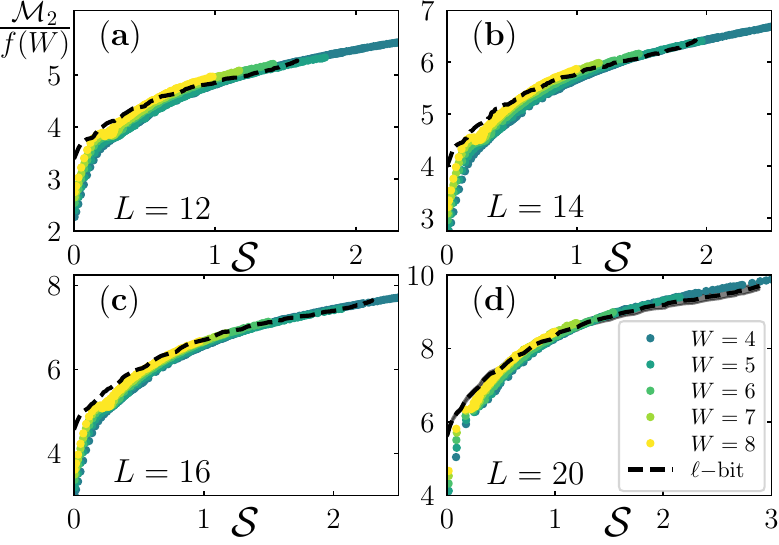}
    \caption{Rescaled SRE ($\mathcal{M}_2/f(W)$) versus entanglement entropy ($\mathcal{S}$) for the initial state $|\Psi_R\rangle$ and (a) $L=12$, (b) $L=14$, (c) $L=16$ and (d) $L=20$. The data collapse improves with increasing $L$, without altering the rescaling parameter $f(W)$. For $L=20$, the data were obtained via Monte-Carlo sampling using $15000$ samples, and the shaded black region indicate the sampling errors. }
    \label{Fig:Collapse}
\end{figure}

\section{Entanglement as the ``internal-clock'' for the nonstabilizerness dynamics}

In the main letter, we show that the entanglement entropy can be seen as an ``internal-clock'' for the SRE growth in the MBL regime, similarly to what occurs with other observables~\cite{Evers23Internal}. Here, however, we show that this result is fully consistent with the $\ell$-bit framework after a proper rescaling by a function $f(W)$ that is independent of the system size $L$. In this section, we explore this idea further by considering the $|\Psi_{Y}\rangle$ and $|\Psi_{R}\rangle$ as the initial state. 

For the former choice, as shown in Fig 3(a) of the main text, both models agree without any need of rescaling by a function $f(W)$. The situation differs if we set $|\Psi_{R}\rangle$ as the initial state. In this case, in order to achieve the collapse between the curves, we minimize the distance between the results obtained for the TFIM with respect to the one obtained for the $\hat{\mathcal{H}}_{\mathrm{\ell-bit}}$. As shown in Fig~\ref{Fig:Collapse}, the quality of the collapse improves as we increase the system size. Most importantly, the rescaling parameter $f(W)$ is independent of $L$, as depicted in the inset of Fig 3(b) of the main text. Hence, our phenomenological description of the SRE growth based on the integrals of motion is fully consistent with the more intricate structure of microscopic Hamiltonians.

\end{document}